\documentclass[a4paper,fleqn,usenatbib]{mnras}

\usepackage[T1]{fontenc}
\usepackage{ae,aecompl}

\usepackage{graphicx}	
\usepackage{amsmath}	
\usepackage{amssymb}	

\newcommand{\kms}{{\rm km\;s$^{-1}$}}

\newcommand{\halpha}{{H$\alpha$}}
\graphicspath{{./figs/}} 

\title[How materials supply to a filament]{How eruptions of a small filament feed materials to a nearby larger-scaled filament}

\author[Wei H. et al.]{
H. Wei, 
Z. Huang,\thanks{E-mail: z.huang@sdu.edu.cn}
Z. Hou,
Y. Qi,
H. Fu,
B. Li
and L. Xia
\\
$^1$Shandong Provincial Key Laboratory of Optical Astronomy and Solar-Terrestrial Environment, Institute of Space Sciences,\\ Shandong University, Weihai, 264209 Shandong, China.\\
$^2$School of Space Sciences and Physics, Shandong University, Weihai, 264209 Shandong, China.
}

\date{Accepted XXX. Received YYY; in original form ZZZ}

\pubyear{2020}

\begin{document}
\label{firstpage}
\pagerange{\pageref{firstpage}--\pageref{lastpage}}
\maketitle

\begin{abstract}
As one of the most common features in the solar atmosphere,
filaments are significant not only in the solar physics but also in the stellar and laboratory plasma physics.
With the New Vacuum Solar Telescope and the Solar Dynamics Observatory,  here we report on multi-wavelength observations of eruptions of a small (30\arcsec) filament (SF) and its consequences while interacting with the ambient magnetic features including a large (300\arcsec) filament (LF).
The eruptions of the SF drive a two-side-loop jet that is a result of magnetic reconnection between the SF threads and an over-lying magnetic channel.
As a consequence of the eruption, the heating in the footpoints of the SF destabilises the barbs of the LF rooted nearby.
Supersonic chromospheric plasma flows along the barbs of the LF are then observed in the \halpha\ passband and they apparently feed materials to the LF.
We suggest they are shock-driven plasma flows or chromospheric evaporations,
which both can be the consequences of the heating in the chromosphere by nonthermal particles generated in the magnetic reconnection associated with the two-side-loop jet.
Our observations demonstrate that the destabilisation in the vicinity of the footpoints of a barb can drive chromospheric plasma feeding to the filament.

\end{abstract}




\section{Introduction}
\label{sect_intro}
Solar filaments (named prominences while they are appearing near and extending above the solar limb) are one of the most common features in the solar atmosphere.
They are dense, cool and partially-ionized plasma structures that anchor to the photosphere and extend outwards into the Sun's hot corona.
They can be clearly seen in the chromospheric emissions, such as \halpha\ passband.
Their maintenance in and interaction with the fully-ionized corona can be of applications in the laboratory plasma.
They might undergo instability and eruption that can develop into solar flares and/or coronal mass ejections\,\citep{2006AdSpR..38.1887W,2011LRSP....8....1C,2011LRSP....8....6S,2014LRSP...11....1P,2014A&A...566A.148H,2017LRSP...14....2B}, thus are of significance in the space weather.
Although solar filaments have been intensively studied for decades, many puzzles remain unsolved because of the complexity of these phenomena\,\citep{2014LRSP...11....1P,2018LRSP...15....7G}. 
Observations with better and better resolution actually add more and more puzzles surrounding the phenomena\,\citep{2018LRSP...15....7G}, because of their high variability that each individual can show different dynamics especially in fine scale at the limit of the instrumental resolution\,\citep{2014LRSP...11....1P}.

\par
As a critical puzzle in the physics of solar filaments, their formation and maintenance have attracted attentions of numerous studies.
It has been well established that solar filaments are formed in filament channels along polarity inversion lines\,\citep[see the reviews by e.g.][]{1998SoPh..182..107M,2010SSRv..151..333M,2014LRSP...11....1P,2018LRSP...15....7G}.
Filament channels are regions where the chromospheric fibrils are aligned with the polarity boundary.
They are believed to be the prerequisite for filament formation\,\citep[e.g.][]{1997ApJ...479..448G,2009ApJ...697..913O,2016A&A...589A..31B}. 
They are voids before filament materials filled in and thus not visible in the remote-sensing images\,\citep[see][and references therein]{1998SoPh..182..107M}.
The magnetic structures of filament channels are believed to be twisted or sheared arcades and thus provide dips to support filament plasmas\,\citep{1989ApJ...343..971V,2000ApJ...539..954D,2001ApJ...558..888G,2004ApJ...612..519V,2007ApJ...666.1284W,2010ApJ...721..901S,2014ApJ...786L..16J,2015ApJS..219...17Y}.
Therefore, there are two critical aspects in understanding the formation of solar filaments.
How is the magnetic structure of a filament channel formed?
How are the cool plasmas transferred into there?

\par
Many efforts have been put on understanding the formation of solar filaments\,\citep[see the recent reviews by][]{2014LRSP...11....1P,2018LRSP...15....7G}.
How a solar filament is formed can be different from case to case (even different from time to time in a particular case), but some basic processes usually occur, including emergence of twisted flux tubes\,\citep[see e.g.][]{2018SoPh..293...93C,2019ApJ...884...45L}, shearing motion parallel to the polarity inversion line\,\citep[see e.g.][]{2001ApJ...560..476C,2015ApJS..219...17Y,2018ApJ...863..192L}, magnetic reconnection among pre-existed (and newly-emerged) magnetic structures\,\citep[see e.g.][]{2016ApJ...816...41Y,2016SoPh..291.2373Z,2017ApJ...839..128W,2017ApJ...840L..23X,2018SoPh..293...93C}, magnetic cancellation at the polarity boundary\,\citep[see e.g.][]{1985AuJPh..38..929M,1990LNP...363....1M,1997ApJ...479..448G,2000SoPh..195..333C,2001ApJ...560..456W,2003ApJ...584.1084C,2013ApJ...763...97W}.
These processes can link to the formation of and/or material supplement to a filament channel.

\par
How materials are supplied to a filament channel has been a hot topic in the community, especially in the past decade when more and more advance instruments are accomplished.
The sources of filament materials are different from case to case\,\citep[see][and references therein]{2014LRSP...11....1P}.
Two well-known scenarios include coronal source (i.e. coronal condensation) and chromospheric source\,\citep[i.e. cool plasma injection, see e.g.][]{2006ApJ...637..531K,2017ApJ...836L..11S}.
For any scenario, how the material supplement to a filament channel takes place remains inconclusive.
Here we summarize a few studies on how cool plasmas are transferred from chromosphere to a filament channel. 
\citet{2000SoPh..195..333C} observed a transient flow field in a system of small \halpha\ loops that some merge into a filament, 
and the observed evolution of the chromospheric flows agrees with a scenario that magnetic reconnection is a way of cool plasma injection in a filament as is proposed by such as \citet{1996ApJ...460..530P}.
\citet{2003ApJ...584.1084C} reported that a series of jets and small eruptions were taking place during the formation of the filament.
\citet{2005ApJ...631L..93L} found that chromospheric surge materials were injected from one terminal along the main axis of the filaments or the filament channels and played an important role in the formation of the filaments.
\citet{2016ApJ...831..123Z} and \citet{2017ApJ...836..122Z} found that filaments are maintained by the continuous injection of cold chromospheric plasma along the filament threads.
\citet{2018ApJ...863..180W} observed the formation of a filament via cool materials with a temperature of about $10^4$\,K ejected by a series of chromospheric jets seen in \halpha\ images.
Their observations also suggest that the jets are results of magnetic reconnection between pre-existed magnetic fields and newly-emerged magnetic fields.
Furthermore, \citet{2019MNRAS.488.3794W} reported in another case that the jets forming the filament consist of both cool chromospheic jets and warmer jets seen in He\,{\sc ii} 304\,\AA.
A recent study with numerical experiment\,\citep{2020NatAs.tmp..110Z} found that turbulences in the chromosphere can drive plasma evaporations that feed materials to a filament.

\par
Here, we report on multi-wavelength observations of the dynamics of active-region filaments in \halpha\ and extreme Ultraviolet (EUV) passbands.
In this case eruptions of a small filament and the consequences of its interactions with the ambient magnetic structures result in magnetic reconfiguration in the region and material supplement to a nearby large filament.
In what follows, we give the data description in Section\,\ref{sect:obs}, the results and discussion in Section\,\ref{sect:res} and the conclusions in Section\,\ref{sect:concl}.

\begin{figure*}
\includegraphics[clip,trim=0cm 0cm 0cm 0cm,width=0.75\textwidth]{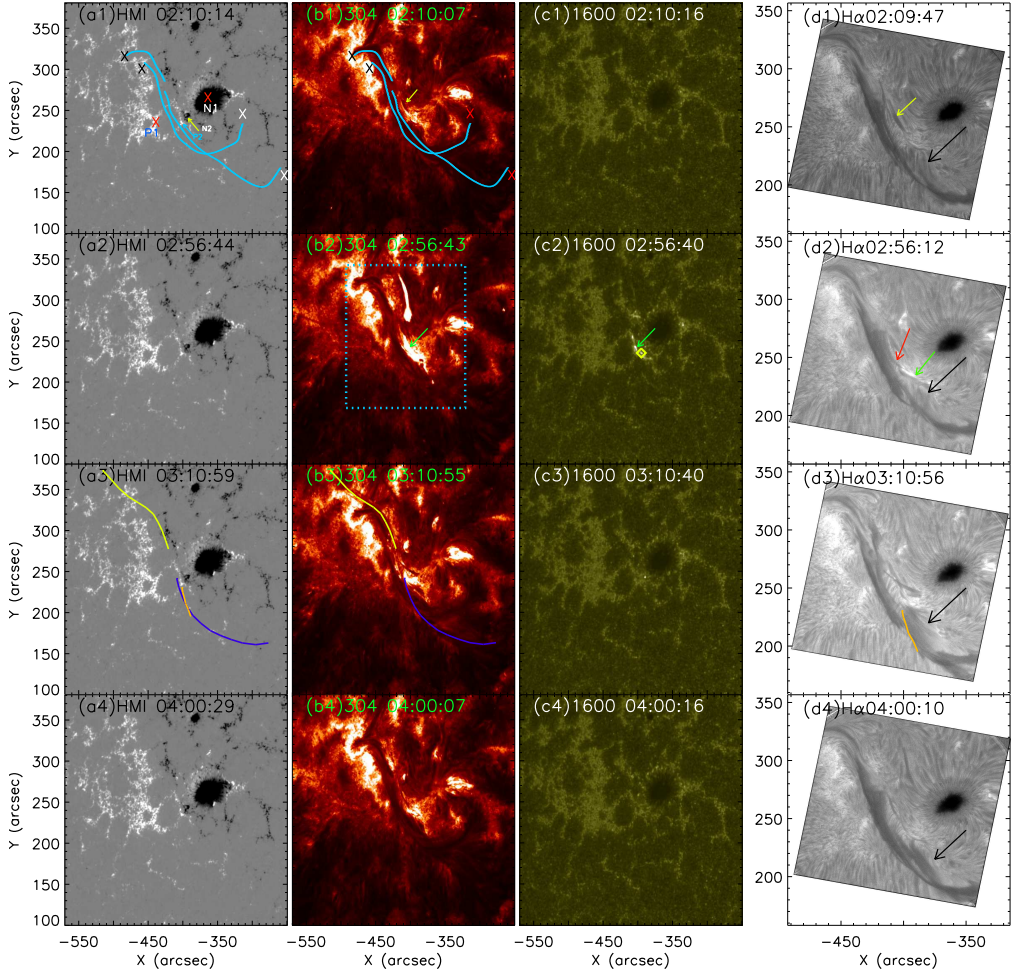}
\caption{The region of interest viewed in the HMI line-of-sight magnetograms (a1--a4), AIA 304\,\AA\ (b1--b4) AIA 1600\,\AA\ (c1--c4) and NVST \halpha\ (d1--d4).
The field-of-view shown in column (d) is marked in panel (b2) as the dotted-line square.
The solid lines in cyan in panels (a1) and (b1) outline a few threads of the LF, and the associated polarities are marked as ``P1'' and ``N1''.
The SF is denoted by yellow arrows in panels (b1) and (d1), and its associated polarities are marked as ``'P2' and ``N2'' in panel (a1).
The crosses in white and black marked in panels (a1) are the visible footpoints of the LF, while the crosses in red are the footpoints of the overlying coronal loops as seen in AIA 171\,\AA\ images.
The red arrow in panel (d2) marks the plasma flow that might be the trigger of the SF eruption.
The green arrows in panels (b2), (c2) and (d2) denote the eruption site.
The black arrows in the 4th column denote the barbs of the LF.
The yellow and blue lines in panels (a3) and (b3) mark the trajectories of the two-side-loop jet.
The orange lines in panels (a3) and (d3) mark the trajectory of the plasma flow along the LF barbs.
The square in yellow in panel (c2) marks the compact brightening that is used to produce the lightcurve shown in Figure\,\ref{fig:st}b.
(An online animation is provided.)
}
\label{fig:mulimgs}
\end{figure*}

\begin{figure}
\includegraphics[clip,trim=0cm 0cm 0cm 0.2cm,width=\linewidth]{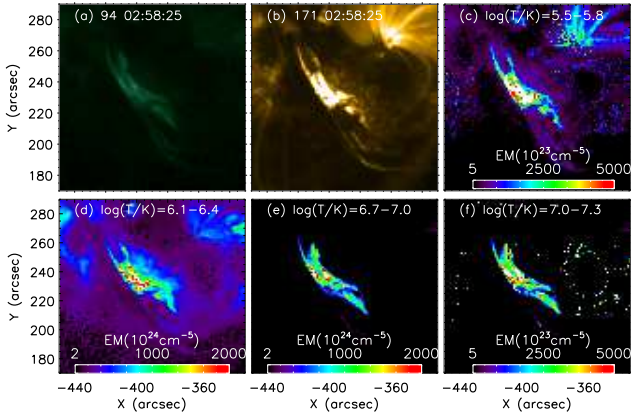}
\caption{The eruption of the SF seen in the AIA 94\,\AA\ and171\,\AA\ passbands and the derived EM maps at various temperature ranges as noted.
(An online animation is provided.)
}
\label{fig:ems}
\end{figure}

\begin{figure}
\includegraphics[clip,trim=0cm 0cm 0cm 0cm,width=\linewidth]{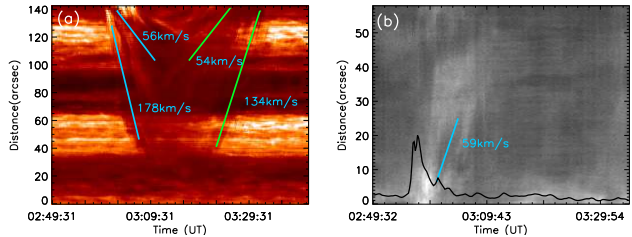}
\caption{(a): Time-distance plots taken from the northward trajectory of the two-side-loop jet (see the yellow lines in Figure\,\ref{fig:mulimgs}a3\&b3) with the AIA 304\,\AA\ observations.
(b): Time-distance plots of the plasma flow along the LF barbs (see the orange lines in Figure\,\ref{fig:mulimgs}a3\&d3) with the NVST \halpha\ observations.
The apparent speeds derived from these plots are marked.
The over-plotted black solid line is the AIA 1600\,\AA\ lightcurve of a compact brightening located in the vicinity the footpoints of the barbs and at the end of the trajectory that used to produced this time-distance map (see the yellow square in Figure\,\ref{fig:mulimgs}c2).
}
\label{fig:st}
\end{figure}

\begin{figure}
\includegraphics[clip,trim=0cm 0cm 0cm 0cm,width=\linewidth]{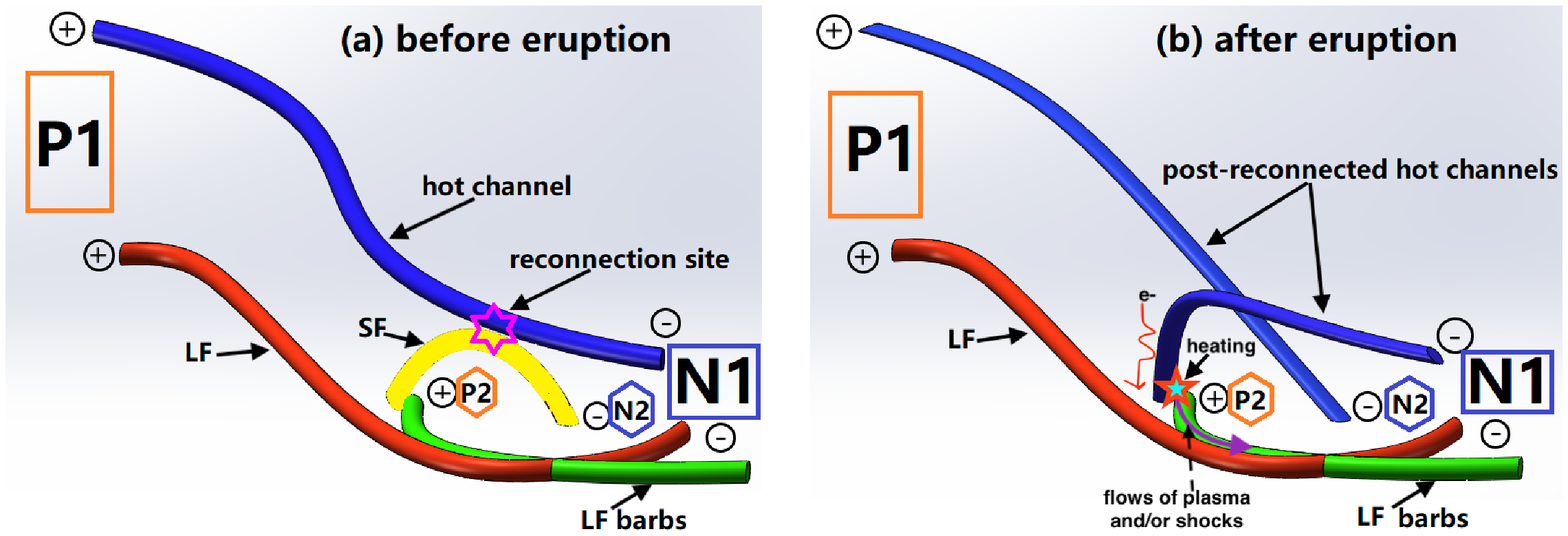}
\caption{A scenario interprets the interaction between the erupted SF and the hot channel and how the materials are fed to the LF. (See the main text for details.)
}
\label{fig:model}
\end{figure}

\section{Observations}
\label{sect:obs}
The observations were taken on March 14 2013 with a target of AR NOAA 11692.
The data were achieved by the ground-based 1-m New Vacuum Solar Telescope\,\citep[NVST,][]{2014RAA....14..705L} of Yunnan Astronomical Observatories of China, 
and the space-born Atmospheric Imaging Assembly\,\citep[AIA,][]{Lemen2012aia} and the Helioseismic and Magnetic Imager\,\citep[HMI,][]{Schou2012hmi} onboard the Solar Dynamics Observatory\,\citep[SDO,][]{Pesnell2012sdo}.

\par
The NVST data include a series of filtergram images taken at the centre of \halpha\ line with a bandpass of 0.25\,\AA.
The data have been fully calibrated and reconstructed\,\citep[see details in][]{2016NewA...49....8X}, and the cadence of the time series is about 12\,seconds.
The stabilisation of the image series was carried out by a fast sub-pixel image registration algorithm\,\citep{Feng2012IEEE,2015RAA....15..569Y}.
The spatial size of a pixel of the \halpha\ images is 0.17\arcsec.

\par
The images taken with the AIA passbands of 1600\,\AA, 304\,\AA, 171\,\AA, 193\,\AA, 211\,\AA, 335\,\AA,131\,\AA\ and 94\,\AA\ are analysed.
The AIA data have a spatial resolution of 1.2\arcsec,
and a cadence of 24\,s for the 1600\,\AA\ passband and 12\,s for the rest.
The analysed HMI line-of-sight magnetograms have a cadence of 45\,s and a spatial resolution of 1.2\arcsec.

\par
The images from different instruments and passbands have been aligned by using multiple reference structures (such as sunspots, bright dots, filament threads, etc.) in passbands with closest representative temperatures.

\section{Results and discussion}
\label{sect:res}
In Figure\,\ref{fig:mulimgs} and the associated animation, we show the evolution of the region of interest as seen in HMI, NVST and AIA.
We can see a large filament (LF) with an apparent length of about 300\arcsec\ extending in the south-north direction, which is clearly shown in \halpha\ and AIA 304\,\AA\ and 171\,\AA\ (see the cyan lines that mark a few filament threads in Figure\,\ref{fig:mulimgs}a1\&b1).
Above the filament, we observe overlying coronal loops in the AIA 171\,\AA\ images (not shown here), which connect the anti-polarity features at both sides of the filament (see the magnetic features at the sides denoted by ``P1'' and ``N1'' in Figure\,\ref{fig:mulimgs}a1).
It indicates that the large filament extends along the polarity inversion line (PIL) between ``P1'' and ``N1''.
Such a magnetic connectivity is also confirmed by magnetic extrapolation carried out for the same active region on the next day\,\citep[][see their Figures 2\&4]{Kawabata_2020}.
At the west side of the LF, a small filament (SF) with an apparent length of about 30\arcsec\ is seen (see the feature denoted by yellow arrows in Figure\,\ref{fig:mulimgs}b1\&d1).
The SF apparently locates along the PIL between the anti-polarity features denoted by ``P2'' and ``N2'' in Figure\,\ref{fig:mulimgs}a1,
and it shows a twisted structure suggesting a sheared field lines\,\citep[see e.g.][]{2000ApJ...539..954D}.
Besides the main bodies of the two filaments, we also see barbs of the LF connecting the region in the vicinity of ``P2'' (see the black arrows in Figure\,\ref{fig:mulimgs}d1).

\par
While tracing the evolution of the region, we observe eruptions of the SF and the consequences of the interactions with the surrounding magnetic structures.
The eruption starts at around 02:56\,UT (see the second row of Figure\,\ref{fig:mulimgs} and the associated animation).
About one minute before the eruption of the SF, we observed a flow with an apparent speed of about 50\,\kms\ (supersonic) injects into the region (appearing as a dark feature in \halpha, see the structure denoted by the red arrow in Figure\,\ref{fig:mulimgs}d2 and can be clearly followed in the associated animation). This injection might plays a role in the destabilization of the SF, but that should require further theoretical study to confirm.

\par
The eruption of the SF shows a complex morphology in the site (see the location denoted by green arrows in Figure\,\ref{fig:mulimgs}b2,c2\&d2) including multiple compact brightenings seen in AIA 1600\,\AA\ observations (see Figure\,\ref{fig:mulimgs}c2).
Due to their compact nature, these brightenings are very likely the sites where the energy was deposited in the chromosphere via nonthermal particles generated in the eruption of the SF.
Many of these brightenings are located in the vicinity of the footpoints of the barbs of the LF.
To investigate the thermal structures of the erupted region, we apply a differential emission measure (DEM) analysis\,\citep{2015ApJ...807..143C,2018ApJ...856L..17S} on the images taken with AIA EUV passbands and obtain the emission measure (EM) of the region.
In Figure\,\ref{fig:ems} and the associated animation, we show zoomed-in views of the erupted region in AIA 94\,\AA\ and 171\,\AA\ channels and EM images at a various temperature ranges.
Besides the multi-thermal nature of the eruption, we can see that the erupted region consists of multiple fine structures in the EM maps, which are consistent with the twisted appearance of the SF.

\par
While the SF is lifted and moving toward the region of the LF, we observed a bi-directional jet (see the trajectories marked by yellow and blue lines in Figure\,\ref{fig:mulimgs}a3\&b3) that has been analysed in detail and interpreted as a two-side-loop jet by \citet{Yang_2019}. 
Small-scaled magnetic cancellations associated with this event have also been reported by \citet{Yang_2019}, suggesting disturbances in the lower atmosphere.
Because we are not focus on the jet, here we show the time-distance plots along the northward trajectory of the jet (see the yellow line in Figure\,\ref{fig:mulimgs}a3\&b3) as seen in AIA 304\,\AA\ only (Figure\,\ref{fig:st}a).
In additional to \citet{Yang_2019}, we would like to point out that the trajectory of the jet does not follow any pre-exited \halpha\ thread of the LF (clearly in the north part, see the yellow line in Figure\,\ref{fig:mulimgs}a3\&b3 and the associated animation of Figure\,\ref{fig:mulimgs}).
This indicates that there is a magnetic channel laid above the LF and it is not visible in the images of \halpha\ and AIA channels.
This ``invisible'' channel might be an empty flux tube or a hot channel that is evidenced by the X-ray images as shown in \citet{Kawabata_2020}.
The jet is most clearly seen with all AIA EUV passbands, and the EM analyses suggest its multi-thermal nature (see Figures\,\ref{fig:mulimgs}\&\ref{fig:ems}).
However, it does not show response in \halpha\ images, again suggesting its origin above the chromosphere.
Moreover, we would also like to mention that the northward jet flow (yellow line in Figure\,\ref{fig:mulimgs}a3\&b3) is reversed about 15 minutes after the initiation of the jet flows and has traveled for about 140\arcsec (see Figure\,\ref{fig:st}a).
After about 03:30\,UT, the hot channel has drained and again disappears from the observations of the AIA EUV passbands.
Thus, the jet does not directly feed materials into the LF.

\par
We also clearly observe plasma flows along the barbs of the LF following to the eruption of the SF (see the black arrows in Figure\,\ref{fig:mulimgs}d1--d4).
The plasma flows can be clearly seen as dark features in the \halpha\ images, and they are mostly cool plasma as seen in the EM images (see Figure\,\ref{fig:ems} and the associated animation).
The flows apparently follow the compact brightening in the footpoints and start as propagations of bright features (see Figure\,\ref{fig:st}b).
By tracing the evolution of the LF in \halpha\ images, we found that these plasma flows apparently change the geometry of the LF (see the feature denoted by black arrow in Figure\,\ref{fig:mulimgs}d4).
Thus, we conclude that these plasma flows have fed materials to the LF.
The flows have an apparent speed of about 59\,\kms\ measured with the \halpha\ time-distance plot (Figure\,\ref{fig:st}b).
Although we cannot rule out the possibility that such apparent flows are results of dark features drifting into the field-of-view,
it is very likely that they are real plasma flows because of the following darkening of the barbs seen in the \halpha\ images.
Since the speed is much larger than the sound speed in the chromosphere,
if they are real plasma flows they should present shock behaviours and a hint is that the brightenings propagate ahead of the dark features in \halpha\ observations (see the bright structure in Figure\,\ref{fig:st}b that start one to two minutes before the dark one as marked by the blue line).

\par
Based on the observations, we give our interpretation for these processes in Figure\,\ref{fig:model}.
Before the eruption (Figure\,\ref{fig:model}a), the LF includes its main threads (represented by the red line) connecting ``P1'' and ``N1'' and barbs connecting ``P2'' and ``N1''; 
the SF connecting ``P2'' and ``N2'' is represented by the yellow line;
the hot channel laid above the LF is shown as the blue line.
When the SF is destabilised, its threads are lifted up and encounter the hot channel above.
That triggers magnetic reconnection between the hot channel and the SF threads,
which generates the two-side-loop jet flows along the reconfigured hot channels (see two blue lines in Figure\,\ref{fig:model}b).
We are aware of that this reconnection does not change the general configuration of the main filament, which is different from that proposed in \citet{1989ApJ...343..971V}.
We believe that the magnetic reconnection produces nonthermal particles, and some of those travel along the field lines toward the chromosphere at ``P2'' that cause heating in the region.
The heating in the chromosphere of ``P2'' destabilises the footpoints of the barbs of the LF and results in shock flows that drive chromosperic plasma feeding to the LF.
Alternatively, the destabilisation in the footpoints of the barbs can cause turbulences, 
which can generate chromospheric evaporation and that feeds chromospheric plasma to the LF as shown in the simulation of \citet{2020NatAs.tmp..110Z}.

\section{Conclusions}
\label{sect:concl}
In the present, we report on NVST and SDO observations of eruptions of a small filament (SF) with a size of about 30\arcsec\ and its interactions with the other magnetic structures including a large filament with a size of about 300\arcsec\ in the region.
The SF and LF are above different polarity inversion lines that belong to different magnetic patches, but there are barbs of the LF connecting the patch of positive polarity of the SF.
The erupted SF threads encounter a magnetic channel above the LF that is previously invisible in \halpha\ and AIA EUV images.
That produces a two-side-loop jet, which is suggested to be a result of magnetic reconnection between the SF threads and the magnetic channel.
As a consequence of the eruption, plasma flows are seen in the barbs of the LF that connect the positive polarity of the SF, and these flows appear to feed materials to the LF.
We interpret that the materials are fed to the LF via shock-driven plasma flows or chromospheric evaporations,
which both can be the consequences of destabilisation in the chromospheric footpoint region from the heating by nonthermal particles generated in the magnetic reconnection associated with the two-side-loop jet.
Our observations demonstrate that the destabilisation in the footpoints of a barb can indeed drive chromospheric plasma feeding to the filament.

\section*{Acknowledgements}
We are grateful to the anonymous referee for the constructive and helpful comments.
This research is supported by the Strategic Priority Program of CAS (XDA15017300), National Natural Science Foundation of China (U1831112, 41974201,  41627806, 11761141002, 41674172),
and the Young Scholar Program of Shandong University, Weihai (2017WHWLJH07).
We would like to thank the NVST operation team for preparation of the data.
Courtesy of NASA/SDO, the AIA and HMI teams and JSOC.

\section*{Data Availability}
The data analysed in this study can be freely requested from the official websites at \textit{http://fso.ynao.ac.cn} (NVST) and \textit{http://jsoc.stanford.edu} (SDO) using the information given in the main text.

\bibliographystyle{mnras}
\bibliography{bibliography}
\bsp	
\label{lastpage}
\end{document}